\documentclass[12pt,preprint]{aastex}

\newcommand{\degree}{\hbox{$^{\circ}$}}

\slugcomment{to appear in ApJ}

\shorttitle{IRAS 16594$-$4656 and IRAS 07027$-$7934}
\shortauthors{Garc\'{\i}a-Hern\'andez et al.}

\begin{document}


\title{Revealing the mid-infrared emission structure of\\ 
IRAS 16594$-$4656 and IRAS 07027$-$7934\footnote{Based on observations
collected at the European Southern Observatory (La Silla, Chile), on
observations made with ISO, an ESA project with  instruments funded by ESA
Member States (especially the PI countries:  France, Germany, the Netherlands
and the United Kingdom) with the  participation of ISAS and NASA, and on
observations made with the NASA/ESA  Hubble Space Telescope, obtained from the
data Archive at the Space Telescope  Science Institute, which is operated by
the Association of Universities for Research in Astronomy, Inc., under NASA
contract NAS5-26555}}


\author{D. A. Garc\'{\i}a-Hern\'andez\altaffilmark{1,2}, A.
Manchado\altaffilmark{1,3},P. Garc\'{\i}a-Lario\altaffilmark{2}, A.
Ben\'{\i}tez Ca\~{n}ete\altaffilmark{4}, J. A. Acosta-Pulido\altaffilmark{1},
and A. M. P\'erez Garc\'{\i}a\altaffilmark{1}}

\altaffiltext{1}{Instituto de Astrof\'{\i}sica de Canarias, E-38205  La Laguna,
Tenerife, Spain; agarcia@iac.es, amt@iac.es, jap@iac.es, apg@iac.es}
\altaffiltext{2}{ISO Data Centre, European Space Astronomy Centre, ESA.
Villafranca del Castillo.  Apartado de Correos 50727. E-28080 Madrid, Spain;
Anibal.Garcia@sciops.esa.int, Pedro.Garcia-Lario@sciops.esa.int}
\altaffiltext{3}{Consejo Superior de Investigaciones Cient\'{\i}ficas, CSIC}
\altaffiltext{4}{Departamento de Astrof\'{\i}sica, Universidad de La Laguna, 
E-38205 La Laguna, Tenerife, Spain}



\begin{abstract}
TIMMI2 diffraction--limited mid-infrared images of a multipolar 
proto-planetary nebula IRAS 16594$-$4656 and a young [WC] elliptical  planetary
nebula IRAS 07027$-$7934 are presented. Their dust shells are for the first
time resolved (only marginally in the case of IRAS 07027$-$7934) by applying
the Lucy-Richardson deconvolution algorithm to the data, taken under
exceptionally good seeing conditions ($\leq$0.5\arcsec). IRAS 16594$-$4656
exhibits a two-peaked morphology at 8.6, 11.5 and 11.7 $\mu$m which is mainly
attributed to emission from PAHs. Our observations suggest that the
central star is surrounded by a toroidal structure observed edge-on 
with a radius of 0.4$\arcsec$ ($\sim$640 AU at an assumed distance of
1.6 kpc) with its polar axis at P.A.$\sim$80\degree, coincident with the
orientation defined by only one of the bipolar outflows identified in
the HST optical images. We suggest that the material expelled from the central
source is currently being collimated in this direction and that the
multiple outflow formation has not been coeval. IRAS 07027$-$7934 shows a
bright, marginally extended emission  (FWHM=0.3$\arcsec$) in the mid-infrared 
with a slightly elongated shape along the N-S direction, consistent with the
morphology detected by HST in the near-infrared. The mid-infrared emission is
interpreted as the result of the combined contribution of small,  highly
ionized PAHs and relatively hot dust continuum. We propose that IRAS
07027$-$7934 may have recently experienced a thermal pulse (likely at
the end of the AGB) which has produced a radical change in the chemistry of
its central star.
\end{abstract}




\section{Introduction}
Planetary nebulae (PNe) are the result of the evolution of low- to
intermediate-mass stars (0.8--8 M$_{\odot}$). These stars experience a phase of
extreme mass loss during the previous asymptotic giant branch (AGB) that causes
the ejection of the stellar envelope. When this mass loss ceases the AGB phase
ends and the star evolves into a short-lived evolutionary stage called the
`post-AGB' or `proto-PN' (PPN) phase just before the star becomes a PN. At
present, the formation of axisymmetric structures in PNe (ranging from
elliptical to bipolar) is believed to be completed by the end of the AGB phase
(Balick \& Frank 2002; Van Winckel 2003). But unlike for PNe, the study of PPNe
is more difficult since their central stars (CSs) are usually too cool to
photoionize the gas. Therefore, we cannot study the formation of axisymmetric
morphologies in PPNe by mapping the ionized gas. We must use alternative
techniques based on the analysis of: (i) the light scattered by the surrounding
dust at optical wavelenghts; (ii) the neutral molecular gas in the envelope in
the near-infrared (H$_2$), submilimeter (e.g. CO) or radio domain (e.g OH, SiO,
H$_{2}$O, CO); and (iii) the dust emission emerging at mid- to far-infrared
wavelengths.

PPNe generally show a double-peaked spectral energy distribution (SED) (Kwok
1993; Volk \& Kwok 1989; van der Veen, Habing \& Geballe 1989) with the
photospheric emission coming from the central  star dominating in the optical
range and a strong infrared excess  indicating the presence of a cool detached
envelope (T$_{d}$$\sim$150--300 K). This strong infrared excess is produced by
the thermal emission of the dust  present in their circumstellar shells
previously expelled during the  AGB phase. The processes that lead to a wide
variety of different morphologies observed in PNe  (e.g. Manchado et al. 2000)
are, however,  still unknown. Several mechanisms have been proposed: the
interaction of stellar winds (e.g. Mellema 1993), binary systems as central
stars (e.g. Bond \& Livio 1990; Morris 1987), non-radial pulsations (e.g. Soker
\& Harpar 1992) or the influence of magnetic fields (e.g. Pascoli 1992; Soker
\& Harpar 1992; Garc\'{\i}a-Segura et al. 1999). To establish  which one(s) of
the above is the dominant process, it is essential to study  these morphologies
as early as possible after the departure from the spherical symmetry takes
place, that is, in the PPN phase. Only recently, with  the help of high spatial
resolution observations it has been possible to study the intrinsic
axisymmetric nature of the dust shells around a few compact PPNe at subarcsec
level (Meixner et al. 1997; Meixner et al. 1999; Ueta et al. 2001).

In this paper, we present for the first time mid-infrared images (8--13 $\mu$m)
at subarcsec level of a PPN IRAS 16594$-$4656 (hereafter I16594) and of a very
young [WC] PN IRAS 07027$-$7934 (hereafter I07027) with the aim of mapping the
dust  emission originated in the innermost regions of their circumstellar dust
shells. The observations made in the mid-infrared are presented in 
Sect. 2 while the data reduction process is described in Sect. 3.  We show  the
results obtained in Sect. 4, which are later discussed in  Sect. 5. The main
conclusions derived from our analysis are given in  Sect. 6.

\section{Mid-infrared Observations}
The observations were carried out on 2001 October 9 and 10, using the imaging
mode of TIMMI2 (Reimann et al. 2000; K\"aufl et al. 2003) attached to the ESO
3.6m telescope  (La Silla, Chile). TIMMI2 has an array of 320 $\times$ 240
pixels with a pixel scale of 0.2$\arcsec$  $\times$ 0.2$\arcsec$ resulting in a
field of view  of 64$\arcsec$ $\times$ 48$\arcsec$.  The observational
conditions were very good (photometric and with a stable seeing of around
0.5$\arcsec$) and, thus, we could obtain mid-infrared images (at 8.6 $\mu$m 
[N1-filter], 11.5 $\mu$m [N11.9-filter] and 11.7 $\mu$m [SiC-filter]) of I16594
and I07027  at the diffraction limit of the telescope. The standard
nodding/chopping observational technique was used in order to cancel the
thermal emission from the atmosphere and from the telescope. An on-chip
nodding/chopping throw of 15" along the north-south  direction was selected.
Due to the short integration times required to avoid saturation in the
mid-infrared, each image is a combination of a large number of individual
sub-images $\sim$50--100, each one with an integration time between 18 and 40
ms and  a chopping frequency around 6 Hz. Total on-source integration times
were typically of $\sim$2 minutes. The mid-infrared photometric standard stars
HD 29291, HD 156277, HD 196171 and HD 6805 (Doublier et al. 2004) were also
observed at different air masses every night to determine the photometric flux
conversion from ADUs (Analog to Digital Units) to Jy and to measure the
telescope point spread function (PSF). 

\section{Data Reduction}
The data reduction process includes bad pixel correction and the combination of
all the images into one single image per filter using standard tasks in
IRAF\footnote{The Image Reduction and Analysis Facility software package (IRAF)
is distributed by the National Optical Astronomy Observatories, which is
operated by the Association of Universities for Research in Astronomy, Inc.,
under cooperative agreement with the National Science Foundation}.  The flux
calibration was made using the conversion factors derived from the observation
of standard stars at different air masses. The variation of these conversion
factors with air mass was slightly different during the two nights of
observation due to the different atmospheric conditions. On October 9 this
variation was very small ($<$10\%), so  a single averaged conversion factor was
used for all the observations performed during the night. However, on October
10 we found larger variations ($\sim$30\%). Thus, for I07027 we took the
conversion factors derived from the observations of the standard star HD
156277, observed closer in time and at a similar air mass. For each filter the
size and  morphology of the target stars in our programme was compared with a
mean PSF derived from the observation of the standard stars used for the flux
calibration.  The object name, observing date, filters, central wavelength and
width of the different filters used, total on-source integration time, object
size, PSF size, integrated and peak fluxes, are listed in Table 1. From the
internal consistency of the measurements made on the standard stars we estimate
that the photometric uncertainty of our observations is of the order of
$\sim$10\%.

The observed PSFs are dominated by diffraction effects. Thus, the
Lucy-Richardson deconvolution algorithm as implemented in IRAF (task {\sc
lucy}) was used in order to  recover the actual emission structure of the
targets in each filter and remove the effects induced by the telescope PSF,
which was found to be very stable. In order to study the goodness of the
deconvolution process the standard stars observed for flux calibration purposes
were also deconvolved with the same PSF used for the target stars. This
exercise is useful to confirm that the deconvolution process does not introduce
undesired artifacts. We found that in all cases the deconvolved images of the
standard stars spread over only one or two pixels, showing a  quasi-point-like
brightness distribution. As an example, one of the deconvolved standard stars
is shown together with its corresponding raw image in Figure 1. 

\section{Results}
\subsection{Mid-infrared Morphology of IRAS 16594$-$4656}
Figure 2 shows the raw images of I16594 taken with TIMMI2 (left panel) together
with the images obtained after applying the Lucy-Richardson deconvolution
algorithm (right panel) above mentioned. The PSFs used for the deconvolution
are also shown for comparison. I16594 shows already in the raw images an
extended elongated morphology surrounding a complex inner core emission (with a
FWHM=1.5" or 3 times the PSF FWHM) which is resolved in more detail after
deconvolution. The source center has been determined by averaging the central
coordinates of the elliptical isophotes within 20--40\% of the peak intensity
(this way we avoid any contamination from the core structure). The images
displayed in Figure 2 have been centered at this position.

The deconvolved 8.6, 11.5 and 11.7 $\mu$m images are shown in the right panel
of Figure 2. The overall elliptical shape of the nebula is clear in all three
filters with its major axis oriented along the east-west direction
(P.A.$\sim$80\degree) and extends out to at least 3.5$\arcsec$ $\times$ 
2.1$\arcsec$ at 5\% of the peak intensity (10-$\sigma$ level above the sky
background) at 8.6 $\mu$m. In addition, a conspicuous double-peaked morphology
in the innermost region of the nebulosity is also recovered, suggesting the
presence of an equatorial density enhancement (e.g. a dust torus). The two
detected peaks are oriented approximately along the north-south direction
(P.A.$\sim$$-$10\degree) perpendicular to the axis of symmetry defined by the
outer elliptical emission. The measured separation between the two peaks is
always 0.8$\arcsec$, independent of the filter considered  (see Figure 3). In
addition, the north-peak (P.A.$\sim$$-$10\degree) is about a factor 2 brighter
than the south peak (see Figure 3). This finding is not unique to I16594.
Indeed, Ueta et al.(2001) found a similar asymmetric profile in IRAS
22272$+$5435. The origin of this asymmetric appearance of the dust torus is
still unclear. Ueta et al. (2001) argued that this can be attributed to
asymmetric mass loss and/or an inhomogeneity in the dust distribution. We are
confident that the deconvolved structure is real because a very similar
emission structure is observed in all three filters. Note that a negligible
contribution to the observed flux at 10 $\mu$m is expected from the central
star of I16594 if this has a B7 spectral type as suggested by Van de Steene,
Wood \& van Hoof (2000).  Hrivnak, Kwok \& Su (1999) found that the central
star contributes only 3\% to the total flux detected in the mid-infrared. We
can therefore safely assume that we are just observing the emission structure
of the dust in the shell alone. The mid-infrared morphology seen in the
deconvolved images is thus interpreted as the evidence of the presence of a
dusty toroidal structure with a 0.4$\arcsec$ radius size seen nearly edge-on.
Recently, Ueta et al. (2005) based on polarization data also suggest an
orientation of the dust torus close to edge-on and they indicate that an
inclination angle of roughly 75 $\degree$ with respect to the line of sight is
derived from a 2-D dust emission model. This adds I16594 to the short list of
PPN where a similar dust torus has been resolved at subarcsec scale. 

\subsection{Mid-infrared Morphology of IRAS 07027$-$7934}
The mid-infrared morphology of I07027 is clearly less complex than the one
observed in I16594 as deduced already from the raw images shown in Figure 4 
(left panel). In this case, only a slightly extended and asymmetric source is 
detected in the two images available, which correspond to the filters N1 and 
N11.9 (centered at 8.6 and 11.5 $\mu$m, respectively). This can just be
due to the larger distance to this source (see Section 5.2.2).  Note,
that the low level extension to the west of the peak emission seen in both
filters seems to be a PSF effect, as it is also observed in the standard stars,
although not so prominent.

 The deconvolved images of I07027 in the 8.6 and 11.5 $\mu$m filters are shown
in Figure 4 (right panel). After the deconvolution, a very bright and slightly
elongated, marginally extended (FWHM=0.3$\arcsec$) emission core is recovered
in both filters, oriented along the north-south direction. A similar
orientation is observed in recent HST-NICMOS images of I07027 taken in the
near-infrared (see Section 5.2). Unfortunately, our TIMMI2 data cannot
confirm precisely whether the mid-IR emission peak is exactly coincident with
the near-IR emission peak. We propose that the emission core which is detected
in the mid-infrared must be coincident with the location of the central
star, as the peak emission is observed exactly at the centre of the
slight extended emission.

\section{Discussion}
\subsection{IRAS 16594$-$4656}
\subsubsection{IRAS 16594$-$4656 in the literature}
I16594 (=GLMP 507) was first identified as a PPN candidate on the basis of its
IRAS colors by Volk \& Kwok (1989) and van der Veen, Habing \& Geballe (1989).
It shows a double-peaked spectral energy distribution  dominated by a strong
mid- to far- infrared dust emission component which is much brighter than the
peak in the near-infrared (Van de Steene, van Hoof \& Wood 2000). The first
indication of the C-rich chemistry of I16594 was the detection of CO molecular
emission in its envelope (with V$_{exp}$$\sim$16 km s$^{-1}$) by Loup et al.
(1990), and the non-detection of OH maser emission (te Lintel Hekkert et al.
1991). More recently, Garc\'{\i}a-Lario et al. (1999) studied the ISO spectrum
of this source and confirmed this classification based on the detection of the
characteristic IR emission features generally attributed to PAHs (at 3.3, 6.2,
7.7, 8.6 and 11.3 $\mu$m) together with relatively strong features at 12.6 and
13.4 $\mu$m which indicates a high degree of hydrogenation in these PAHs. The
ISO spectrum also reveals the presence of strong 21, 26 and 30 $\mu$m dust
emission features (see Figure 5), adding I16594 to a short list of known PPNe
displaying this set of still unidentified features.

The optical spectrum of I16594 shows only the hydrogen Balmer emission lines
over an extremely reddened stellar continuum (E$_{B-V}$=1.8, Van de Steene \&
van Hoof 2003) consistent with a B7 spectral type if dereddened. HST optical
images show the presence of a bright central star surrounded by a multiple-axis
bipolar nebulosity (seen in scattered light) with a complex morphology at some
intermediate viewing angle (see Figure 6). The size of this optical nebulosity
is 6.3$\arcsec$ $\times$ 3.3$\arcsec$ at 3$\sigma_{sky}$ level (Hrivnak, Kwok
\& Su 1999). 

 In the literature there are several indications of the presence of a
circumstellar disc or a torus (an equatorial density enhancement) around
I16594. The highly collimated structure seen in the HST optical images and the
non-detected radio-continuum emission ($<$10 $mJy$) by Van de Steene \&
Pottasch (1993) suggest that the emission lines observed in the optical
spectrum are the result of shock excitation produced by a fast bipolar wind
from the central source in interaction with the slow AGB wind. In agreement
with this hypothesis Garc\'{\i}a-Hern\'andez et al. (2002) reported the
detection of H$_2$ shock-excited emission in I16594, later confirmed by Van de
Steene \& van Hoof (2003) through a more detailed analysis of the H$_2$
spectrum. They postulate that the H$_2$ emission originates mainly where the
stellar wind is funnelled through a circumstellar disc or torus. More recently,
Hrivnak, Kelly \& Su (2004) presented HST-NICMOS near-infrared images of
I16594 which show that this emission is originated in regions where shocks must
be taking place. Polarization measurements originally taken by Su et al. (2003)
and later analyzed by Ueta et al. (2005), who presented a PSF subtracted map of
the polarized light, suggest the presence of an equatorial  enhancement in
I16594 as well. However, Van de Steene, van Hoof \& Wood (2000) failed to
detect any extended emission in their N-band TIMMI images of I16594 in a
previous attempt to search for mid-infrared emission coming from this torus,
but they observed the source  with a lower spatial resolution (pixel scale of 
0.66$\arcsec$), and under poor weather conditions.

\subsubsection{A Dusty Toroidal Structure around IRAS 16594$-$4656}
There exists more than a dozen PPN shells that have been resolved in
the mid-infrared so far. However, only a few of them show some structure at
mid-infrared wavelengths. Meixner et al. (1999) found two different classes of
mid-infrared morphologies. They distinguish those sources with a mid-infrared
core/elliptical structure from those with a toroidal one and they argue that
this morphological dichotomy is due to a difference in optical depth. In their
sample there are only 4 out of 6 toroidal PPN/PNe in which the central dust
torus is well resolved in two emission peaks. This work adds I16594 to this
short list. In Table 2 we list the few known toroidal PPN/PNe sample together
with some of their main observational characteristics, such as the spectral
type of the central star,  C/O ratio, evolutionary classification, optical
morphology, and the list of  mid-infrared dust emission features detected.

An inspection of Table 2 clearly indicates that I16594 is now a
toroidal-PPN with the earliest spectral type known. The other PPNe with
mid-infrared toroidal structures have all F-G spectral types, while IRAS
21282$+$5050 is already a young PN with an O9-type central star. It seems that
many of these sources show a C-rich chemistry (indicated by the presence of PAH
emission features) but the number of objects considered is still small and the
statistics are very poor. It is interesting to remark the fact that all 
mid-infrared toroidal-PPN/PNe have bipolar/multipolar optical morphologies
where the central star is clearly seen. In contrast, the central star is rarely
seen in the mid-infrared core/elliptical class sources described by Meixner et
al. (1999) and almost all of them display bipolar morphologies in the optical.
In addition, the mid-infrared core/elliptical sources are typically O-rich and
show deep silicate absorption features at 9.8 $\mu$m in their mid-infrared
spectra, indicating that they may be optically thick at mid-infrared
wavelengths (Meixner et al. 1999). The different optical morphology (with or
without a visible central star) and the apparent differences in dust properties
(optical thickness in the mid-infrared) suggest that mid-infrared toroidal PPNe
might be surrounded by a dust torus which is optically thin at mid-infrared
wavelengths and, thus, not able to obscure the central star in the optical
domain, while, in contrast, mid-infrared core/elliptical PPNe would be
surrounded by an optically thick dust torus/disk which would completely obscure
the central star in the optical (Meixner et al. 1999, 2002; Ueta et al. 2000,
2003). 

Our deconvolved mid-infrared images of I16594, of much better quality
than those previously reported by Van de Steene, van Hoof \& Wood
(2000), reveal directly for the first time the presence of an 
optically thin dusty toroidal structure with a radius of 0.4$\arcsec$.
Unfortunately, the distance determinations to I16594 are quite uncertain
and, thus, a direct transformation of this observed size into an
absolute physical value is not straightforward. Estimations based on the
observed reddening are hampered by the fact that the overall extinction is
always a combination of interstellar and circumstellar reddening. And in the
case of I16594 there seems to be a considerable contribution from the
circumstellar component. Calculations made by  Van de Steene \& van Hoof (2003)
based on the intrinsic colors expected for a B7 central star in the optical and
in the near-infrared suggest a total extinction of A$_{V}$=7.5 mag with
R$_{V}$=4.2. With this value for the extinction and the flux calibration from
the Kurucz model a distance of (2.2$\pm$0.4) L$_4$$^{1/2}$ kpc is obtained,
where L$_4$ is in units of 10$^4$ L$_{\odot}$. We have tried to derive our own
distance  estimate to I16594 based on the analysis of the overall SED, from the
optical to the far-infrared. For this we put together the IRAS fluxes at 12,
25, 60 and 100 $\mu$m, the near-infrared JHKL magnitudes from \citet{gl97} and
the BVRI magnitudes from \citet{hr99}. The observed BVRI and JHKL fluxes were
corrected for extinction using the total extinction of A$_{V}$=7.5 mag
determined by \citet{vv03} and the extinction law from \citet{ca89}. Then, a 
distance-dependent luminosity was obtained by integrating the observed flux at
all wavelengths and extrapolating the IRAS fluxes to the infinite following
Myers et al. (1987). This way, a distance of 2.1 L$_4$$^{1/2}$ kpc is obtained,
in very good agreement with the previous determination by Van de Steene \& van
Hoof (2003). Assuming a luminosity of 6,000 L$_{\odot}$, which is the
theoretical luminosity expected for a post-AGB star with a core mass of 0.60
M$_{\odot}$ (Sch\"onberner 1987), a distance of 1.6 kpc to I16594 is derived,
value that will be adopted in the following discussion. The value of 0.60
M$_{\odot}$ is chosen for the mass of the core because the mass distribution of
planetary nebulae central stars is strongly peaked at this value (Stasynska,
Gorny \& Tylenda 1997).

At a distance of 1.6 kpc, the extended emission detected in our deconvolved
mid-infrared images of I16594 would correspond to a dusty toroidal structure
with a radius of $\sim$640 AU. Assuming that the CO emission detected towards
I16594 is a good tracer of the dusty torus structure and considering the CO
expansion velocity of 16 km s$^{-1}$ measured by Loup et al. (1990), a
dynamical age of the dusty torus structure of $\sim$190 yr can be estimated.
This dynamical age is quite consistent with a source which has left the AGB
very recently.

\subsubsection{Dust Temperature}
The radiation transfer equation in the interior of a dust cloud adopts a simple
form when the energy source is a single exciting star under the optically thin
approximation and assuming thermal equilibrium. Under these conditions, the
mean color temperature of the dust can be obtained from a simple equation (see
e.g. Evans 1980) which relates the measured fluxes S$_{\nu1,2}$ at two
different wavelenghts $\lambda_{1,2}$ and the dust emissivity index (which
depends on the assumed dust model), assuming a homogeneous dust distribution
throughout the cloud. It should be noted that the assumption of the central
star as the only source of energy for dust heating is appropiate for I16594
because direct stellar ratiation is the dominant heating source for the
circumstellar dust grains in the shell. In particular, we have investigated
whether dust heating due to line emission could also contribute to the observed
emission and found that this effect is negligible, as the value of the Infrared
Excess(IRE) for I16594, defined as the ratio between the observed total far
infrared flux and the expected far infrared flux due to absorption by dust of
Ly $\alpha$ photons (see e.g. Zijlstra et al.1989), is $\sim$ 400. 

In principle, one could construct color temperature maps for the dust from the
analysis of the 8.6 and 11.5 $\mu$m TIMMI2 images of any given source, as long
as these bands are representative of the dust continuum emission.
Unfortunately, in the case of I16594, the 8.6 and 11.5 $\mu$m emission is
strongly affected by the PAH emission features which are clearly visible in the
ISO spectrum (see Figure 5). Thus, the temperature values derived this way are
not expected to represent realistic estimations of any physical temperature in
the shell. The same problem is found if we try to derive the dust temperature
from the IRAS photometry at 12 and 25 $\mu$m, since both filters are also
strongly affected by the presence of dust features, as ISO spectroscopy
reveals. This is confirmed by the strongly different mean dust temperatures
T$_{8.6/11.5}$ of 227 K and T$_{12/25}$ of 129 K, derived (assuming a
dust emissivity index of 1) from our mid-infrared data and from the IRAS
photometry at 12 and 25 $\mu$m, respectively.

A more reliable dust temperature can be directly estimated from the
observed size of 0.4$\arcsec$ for the inner radius of the dusty torus assuming
that this is the equilibrium radius for the  bulk of the dust emitting at
mid-infrared wavelengths. Based on the formula worked out by Scoville \& Kwan
(1976), this can be calculated using the equation:  \begin{equation} T_{d} =
1.64f^{-1/5}r_{eq}^{-2/5}L_{*}^{1/5} \end{equation} where \textit{T$_{d}$} is
the dust temperature in K, \textit{f} is the emissivity of the dust, r$_{eq}$
is the equilibrium radius in $pc$, and  \textit{L$_{*}$} is the source
luminosity in L$_{\odot}$.

We decided to use for our calculations a basic dust model composed by 
hydrogenated amorphous carbon grains (HACs; type BE of Colangeli et al. 1995),
whose emissivity index is $\sim$1. The selection of this dust model to
reproduce the dust continuum emission observed in I16594 is justified by the
presence of highly hydrogenated PAHs in the ISO spectrum (Garc\'{\i}a-Lario et
al. 1999). In order to calculate the dust emissivity, the mass
extinction coefficient value for hydrogenated amorphous carbon was taken from
appendix A of Colangeli et al. (1995) at the central wavelength between the two
filters. Then, a typical grain density of 1.81 g cm$^{-3}$ (Koike, Hasegawa \&
Manabe 1980) was assumed. Finally, this quantity was multiplied by a dust grain
size in the range 0.001-0.1 $\mu$m obtaining a dust emissivity \textit{f}. Note
that a dust grain size of 0.01 $\mu$m is a reasonable mid-range size for
circumstellar carbon dust (e.g. Jura, Balm \& Kahane 1995). The dust
temperature in thermal equilibrium at 0.4$\arcsec$ (or $\sim$640 AU at the
assumed distance of 1.6 kpc) can then be derived using the dust emissivity
\textit{f} and the assumed luminosity of the source (6,000 L$_{\odot}$).
This way, a dust temperature T$_{d}$=237 K is found for a mid-range 
dust grain size of 0.01 $\mu$m. A smaller or a larger dust grain size of 0.001
and 0.1 $\mu$m would imply dust temperatures of 376 and 150 K, respectively.

We are conscious that the assumption of spherical geometry may not be valid for
the circumstellar envelope of I16594 where the emission is clearly asymmetric
and the geometry assumes a toroidal shape, according to our mid-IR images. Note
that adopting a more complex, axysimmetric geometry would esentially translate
into grains being more effectively heated in the biconical opening angle
defined by the dust torus  because of the different local optical depth. In
spite of this, our simple model can be used irrespective of the shell geometry
when applied to dust grains at the inner radius of the shell. The use of more
detailed axysimmetric, multiple grain size models is beyond the scope of this
paper. In addition, the presence of a dust torus close to the star with respect
the spherical case mainly influences the optical and near-infrared radiation. A
large effect on the mid- to far-IR emission is not expected (see e.g. Ueta \&
Meixner 2003). In this sense, spatially unresolved SEDs do not provide any
spatial information necessary to constrain the geometry and inclination angle
of the PPN dusty shells.

\subsubsection{Comparison with ISO data}
Another dust temperature estimate can be derived by fitting one (or more)
blackbodies to the available ISO data by considering fluxes representative of
the underlying continuum at carefully selected wavelengths not affected by any
dust feature. We did this by selecting the ISO fluxes at 6.0, 9.4, 14.3, 18.0,
and 45.0 $\mu$m plus the IRAS fluxes at 60 and 100  $\mu$m. The best fit to the
overall SED is obtained with a combination of two blackbodies (with an
emissivity index of 1) with temperatures of 273 K and 130 K, respectively, as
we can see in Figure 5, where we display the SED of I16594 from 1 to 100
microns together with the two blackbodies. We find that actually the warm
component (at 273 K) dominates in the wavelength range of the N1-filter (at 8.6
$\mu$m) while the cool component (at 130 K) dominates in the N11.9-filter range
(at 11.5 $\mu$m). Overimposed on the continuum emission, strong PAH features
are also clearly contributing to the observed emission. Note that the PAH
emission features observed at the ISO short wavelengths as well as the dust
features at 21, 26 and 30 $\mu$m, the latter extending from 20 to 40 $\mu$m,
are intentionally excluded from the fitting because they are not representative
of the dust continuum emission. In particular, the 30 $\mu$m feature overlaps
with the 26 $\mu$m feature, and even with the 21 $\mu$m feature and the
continuum level is well below the flux detected by ISO at 22--24 microns (see
e.g. the analysis of the similar sources IRAS 20000$+$3239 and HD 56126 shown
in Fig. 10 of Hony, Waters \& Tielens 2002). At present, most of  these
features remain still unidentified, although several possible carriers have
been proposed in  the literature, e.g. fullerenes, TiC, SiC for the 21 $\mu$m
feature  (Garc\'{\i}a-Lario et al. 1999; von Helden et al. 2000; Speck \&
Hofmeister 2003); MgS for the broad 30 $\mu$m feature (Hony, Waters, \& Tielens
2002 and references therein). 

Using the above two dust temperatures we can estimate the size of the dust
grains which are expected to emit in equilibrium at the distance of
0.4$\arcsec$ from the central star which is derived from our mid-IR images.
This is found to correspond to small dust grains with a size of 0.005 $\mu$m in
the case of the warm dust component emitting at 273 K which dominates at 8.6
$\mu$m (comparable to the typical size of small PAH clusters). Dust grains with
the same size emitting at 130 K (note that this cold dust emission
dominates at 11.5 and 11.7 $\mu$m) would need to be located at $\sim$4075 AU
from the central star, which corresponds to a projected $\sim$2.5$\arcsec$ on
the sky at the assumed distance. This is considerably beyond the observed
extension of the inner shell in the mid-infrared. A surface brightness of
$\sim$350 mJy/pixel can be roughly estimated for the continuum emission
expected under these conditions, well above ($\sim$50-$\sigma$) our detection
limit. The fact that we do not detect this extended emission in our images
suggests that the angular size of the region giving rise to the bulk of the hot
dust emission is much smaller than that of the region emitting at 130 K. Note
that, assuming e.g. that the cold dust emission extends homogeneously over the
larger aperture used by ISO, we find that the surface brightness would be just
below the 3-$\sigma$ level of the sky background and, as such, undetectable in
our TIMMI2 images. The similar extension and morphology of the mid-infrared
emission observed at 8.6, 11.5 and 11.7 $\mu$m suggests that the contribution
from PAHs observed in the ISO spectrum must be dominant in our TIMMI2 images,
and that these PAHs may be well mixed with the small, hot dust grains
responsible for the underlying continuum, being mainly distributed along the
torus. 

Considering the information available and the limited spectral coverage,
an alternative scenario which cannot be ruled out completely might be that both
small, hot dust grains and large, cold dust grains could be co-located in the
dust torus. This would be possible if a larger grain size ($\geq$0.1 $\mu$m) is
assumed for the cold dust. Note that, in a non-spherical (torus) distribution
of the dust, the shielding can become very efficient and the density very high
in the outer equatorial regions, where the dust can grow and get colder,
protected both from the radiation from central star and from the ISM UV
radiation field. This would explain the larger size of the cold dust  grains in
the torus. In contrast, small, hot dust grains are expected to dominate in the
inner boundary of the torus. Unfortunately, the spatial resolution of our
images is not enough to resolve the grain size distribution within the torus.

\subsubsection{Collimated outflows in IRAS 16594$-$4656}
I16594 has also been observed by the HST in the optical, through the broad
F606W continuum filter with the Wide Field Planetary Camera (WFPC2) under
proposal 6565 (P.I.: Sun Kwok), and in the near-infrared, through the narrow
F212N (H$_2$) and F215N (H$_2$-continuum) filters with the Near Infrared Camera
and Multi Object Spectrometer (NICMOS) under proposal 9366 (P.I.: Bruce
Hrivnak). In the optical, I16594 shows a flower-shaped morphology where several
petals (or bipolar lobes) can be identified at the opposite sides of the
central star with different orientations, which has been suggested to be a
result of episodic mass  ejection (Hrivnak, Kwok, \& Su 1999). Similar
structures have also been detected in other PPNe (e.g. Hen 3-1475; Riera et al.
2003) and in more evolved PNe (e.g. NGC 6881; Guerrero \& Manchado 1998) and
they have been interpreted as the result of episodic mass loss from a
precessing central source (e.g. Garc\'{\i}a-Segura \& L\'opez 2000). From
the HST optical images (taken from the HST Data Archive) we identify pairs of
elongated structures with at least four different bipolar axes at
P.A.$\sim$34$\degree$, $\sim$54$\degree$, $\sim$84$\degree$ and
$\sim$124$\degree$. 

In Figure 6 we have displayed the contour map of the deconvolved mid-infrared
images of I16594 obtained with TIMMI2 in the N1 and N11.9 filters overlaid on
the optical HST-WFPC2 image taken in the F606W filter. Remarkably, we can see
that the axis of symmetry defined by the mid-infrared emission nicely coincides
with only one of the bipolar axes that can be identified in the optical images,
in particular with that oriented at P.A.$\sim$84$\degree$. If this emission is
a good tracer of the hot dust in the envelope and we accept that this hot dust
must have been recently ejected from the central star we can interpret the
observed spatial distribution in the mid-infrared as the result of the
preferential collimation of the outflow material along this direction in the
most recent past.

Remarkably, the H$_2$ shocked emission detected with HST-NICMOS in the 
near-infrared is also found mainly distributed following the same bipolar axis
(Hrivnak, Kelly \& Su 2004) and nicely coincides with the mid-IR emission seen
in our TIMMI2 images. This is shown in Figure 7, where the H$_2$
continuum-subtracted HST-NICMOS image is shown together with a contour map of
the deconvolved mid-infrared image taken in the N11.9 filter. Note that
the H$_2$ image (at 2.122 $\mu$m) showed in Figure 7 was continuum-subtracted
using the HST-NICMOS image taken in the adjacent continuum at 2.15 $\mu$m (both
images were also taken from the HST Data Archive). Interestingly, we found
that the H$_2$ emission is mainly coming from the walls of the bipolar lobe
oriented at P.A.$\sim$84$\degree$ identified in the HST optical images. In
addition, four additional clumps of much weaker H$_2$ emission are detected at
the end of each of the other two point-symmetric outflows associated to I16594
(Hrivnak, Kelly \& Su 2004). The stronger emission detected along the walls of
this bipolar lobe suggests that the interaction of the fast wind from the
central star with the slowly moving AGB wind is currently taking place
preferentially also along this axis of symmetry. This suggests that the
formation of the multiple outflows observed in I16594 has not been
simultaneous. The rest of bipolar outflows observed at other orientations in
the optical images taken with HST must then be interpreted as the result of
past episodic mass loss ejections. As such, they must contain much cooler dust
grains which are then only detectable in the optical because of their
scattering properties.

\subsection{IRAS 07027$-$7934}
\subsubsection{IRAS 07027$-$7934 in the literature} 
I07027 (=GLMP 170) is a very peculiar young PN. It has a central star that was
classified by Menzies \& Wolstencroft (1990) as of [WC11]-type. At present,
there are only about half a dozen PNe with a central star classified as [WC11].
They all have stellar temperatures between $\sim$28,000 and 35,000 K
(Leuenhagen \& Hamann 1998) and are supposed to be in the earliest observable
phase of its PN evolution, soon after the onset of the ionization in their
circumstellar envelopes. I07027 is also among the brightest IRAS PNe and it has
IRAS colors similar to other young PNe (Zijlstra 2001). The youth of I07027 as
a PN is also  evidenced  by the detection of OH maser emission at 1612 MHz
(Zijlstra et al. 1991), which is usually observed in their precursors, the
OH/IR stars, but very rarely in PNe. The OH emission is single-peaked, which is
interpreted as being detected only coming from the blue side of the shell, as
the consequence of the ionized inner region being optically thick at 1612 MHz.
This is supported by the shift in velocity with respect to the CO emission,
which has also been detected toward this source, and from which an expansion
velocity of 14.5 km s$^{-1}$ is derived (Zijlstra et al. 1991).

The detection of strong PAH features and  crystalline silicates in the ISO
spectrum (Cohen et al. 2002; Peeters et al. 2002) indicates the simultaneous
presence of oxygen and carbon-rich dust in the envelope. Remarkably, all other
[WC] CSPNe observed with ISO show a mixed chemistry as well (Cohen et al. 2002)
but I07027 is the only known [WC] star belonging to the rare group of PNe with
OH maser emission, and therefore it links OH/IR stars with carbon-rich PNe. 

Zijlstra et al. (1991) published an H$\alpha$ image of I07027 taken with the
ESO 3.5m NTT telescope. This image shows a stellar core with non-gaussian wings
extending to a maximum diameter of about 15$\arcsec$, which may be mostly due
to  light scattered by  neutral material and dust grains in the envelope.
Garc\'{\i}a-Hern\'andez et al. (2002) detected H$_2$ fluorescence-excited
emission from this source, in agreement with the round/elliptical H$\alpha$
morphology of the nebula and the temperature of the central star. 

I07027 had never been imaged in the mid-infrared before. Thus, our 
observations are the first attempt to reveal the spatial distribution of the 
warm dust in this peculiar object.

\subsubsection{The Marginally Extended Mid-infrared Core of IRAS 07027$-$7934}
The deconvolved mid-infrared images of I07027 displayed in Figure 4 show a
slightly extended emission at 8.6 and 11.5 $\mu$m. This mid-infrared emission
is only marginally resolved (with a FWHM=0.3$\arcsec$ as compared to the
typical PSF size of FWHM$\leq$0.2$\arcsec$ measured in the deconvolved standard
stars) and is elongated along the north-south direction. Zijlstra et al. (1991)
predicted for this source a radio flux density of 10 $mJy$ assuming
E$_{B-V}$=1.1 and T$_{e}$=10$^4$ K. In addition, by using a plausible radio
brightness temperature of 10$^3$ K they predicted an angular diameter of
$\sim$0.3$\arcsec$ for the ionized region. This size for the ionized region is
consistent with the measured size of the bright mid-infrared core seen in our
deconvolved images of I07027.

Unfortunately, there are no HST images of I07027 available  in the optical but
it has very recently been observed in the near-infrared through the broad F110W
(J-band) and F160W (H-band) continuum filters with NICMOS under proposal 9861
(P.I.: Raghvendra Sahai). In Figure 8 we have displayed the still unpublished
near-infrared HST images of I07027 (taken from the HST Data Archive) together
with the contour levels of the deconvolved mid-infrared image taken by us in
the N11.9 filter. In the F160W filter, I07027 shows a bright extended core
(with FWHM=0.25$\arcsec$), which is slightly elongated along the north-south
direction, in agreement with the mid-infrared structure seen in our deconvolved
TIMMI2 images. This core is surrounded by a fainter elliptical nebulosity
extended along the NW-SE direction with a total size of $\sim$1.6$\arcsec$
$\times$ 2.1$\arcsec$ at 1\% of the peak intensity. The HST image in the F110W
filter shows a slightly less extended emission of $\sim$1.1$\arcsec$ $\times$
1.5$\arcsec$ (at 1\% of the peak intensity) and shows a very similar
morphology. In this case, a central point source is clearly detected which
corresponds very probably to the  central star, which is barely detected in the
F160W image.

Similarly to what we did for I16594 we have also estimated the distance to
I07027. In this case we constructed the SED of I07027 by combining the
available IRAS fluxes at 12, 25, 60 and 100 $\mu$m with the JHKL and BVRI
photometry taken from Garc\'{\i}a-Lario et al. (1997) and Zijlstra et al.
(1991), respectively. The observed fluxes were also corrected for reddening
adopting the extinction law from Cardelli, Clayton \& Mathis (1989) and the 
value of E$_{B-V}$=1.1 derived by Zijlstra et al. (1991) through the 
measurement of nearby stars, with R$_{V}$=3.1. Then, a distance of 4.1
L$_4$$^{1/2}$ kpc is obtained. Note that I07027 is located at a much higher
galactic latitude (b=$-$26\degree) than I16594 (b=$-$3\degree) and at such high
galactic latitudes so much interestellar reddening is unexpected. Thus, we
interpret that the observed reddening E$_{B-V}$=1.1 is mainly circumstellar in
origin. On the other hand, most of the flux is emitted in the infrared where
the effect of the interstellar/circumstellar extinction is mild. This is
probably the reason why a very similar luminosity of 4.2 L$_4$$^{1/2}$ kpc was
obtained by Surendiranath (2002), who derived this value by integrating the
photometric fluxes from 0.36 $\mu$m to 100 $\mu$m, but without introducing any
correction for extinction. Assuming a standard luminosity of 6,000 L$_{\odot}$
for I07027, a distance of 3.2 kpc is derived, in agreement with the distance of
3--5 kpc suggested by Zilstra et al. (1991). At this distance, the core size
would correspond to $\sim$960 AU. 

\subsubsection{Dust Temperature}
As for I16594, stellar light must be the dominant heating source for the
circumstellar dust grains in I07027. This is confirmed by Zijlstra et al.
(1991), who derived an Infrared Excess (IRE) of 93 for this source, indicating
that dust heating by line emission can also be neglected in I07027. Again, we
cannot interpret our mid-infrared observations of I07027 in terms of dust
temperatures in the shell because the ISO spectrum of I07027 (Cohen et al.
2002) shows that the 8.6 and 11.5 $\mu$m filters are also heavily affected by
strong PAH emission features. In particular, the PAH emission features around 8
$\mu$m (at $\sim$7.7 and 8.6 $\mu$m) are much stronger in this case than the
feature located at 11.3 $\mu$m. Thus, the dust temperature values derived would
be unrealistically high. Using IRAS data and assuming a dust emissivity
index of 1, a mean dust temperature T$_{12/25}$ of 148 K is derived, while the
TIMMI2 data gives a T$_{8.6/11.5}$ of $\sim$363 K. The strong differences in
the derived temperatures confirm that the PAH emission is dominating the
emission observed in the mid-infrared. 

In contrast to I16594, the ISO spectrum of I07027 shows much weaker emission
features at 12.6 and 13.4 $\mu$m, which are the signatures of the CH
out-of-plane bending vibrations for hydrogens in positions duo and trio,
respectively (Pauzat, Talbit \& Ellinger 1997) and indicate that the PAH
population in I07027 is largely dehydrogenated. Then, for the modelling of the
dust emitting at mid-IR wavelengths we made the same assumptions as in the case
of I16594 (see Section 5.1.3) but this time we adopted a composition dominated
by dehydrogenated amorphous carbon grains (type ACAR of Colangeli et al. 1995).
Under these assumptions and taking into account the dust equilibrium radius to
be consistent with the 0.3$\arcsec$ (or $\sim$960 AU at the assumed distance of
3.2 kpc) of the shell (which is the radius at $\sim$90\% of the peak intensity)
seen in our mid-infrared deconvolved images we obtain a dust temperature of 219
K for a mid-range dust grain size of 0.01 $\mu$m. For a larger dust grain size
of 0.1 $\mu$m a smaller dust temperature of 138 K is derived while a smaller
dust grain size of 0.001 $\mu$m yields a dust temperature of 347 K. Note that
if the dust grains were located closer to the central star than the
0.3$\arcsec$ derived from our mid-infrared images, the dust temperatures above
derived should then be considered as lower limits.

\subsubsection{Comparison with ISO data}
The validity of the range of possible dust temperatures derived from
our TIMMI2 observations can be further explored by looking at the ISO spectrum
originally published by Cohen et al. (2002). In a similar way as we did for
I16594, the SED can be fitted by a two-component dust continuum with
temperatures of T$_{BB1}$=430 K and T$_{BB2}$=110 K, respectively. The warm
component in this case completely dominates in the wavelength range of our
TIMMI2 observations (where also strong PAH features are found) while the cool
component dominates at longer wavelengths, where crystalline silicate dust
features are also detected on top of the continuum emission.

By forcing the dust equilibrium radius to be consistent with the
0.3$\arcsec$ seen in our mid-IR deconvolved images, we need to assume in this
case a very small grain size of $<$0.001 $\mu$m in order to reproduce the
dust temperature of 430 K derived from the ISO spectrum. This suggests that
the mid-infrared emission at $\sim$430 K must be the result of the combined
contribution of small PAH molecules, located very close to the central star,
and relatively hot dust continuum. In this case, the PAH population must be
subject to a relatively strong UV field, consistent with the narrow features
detected by ISO (in contrast to the broader features observed in I16594).
Actually, the ISO spectrum of I07027 shows that the PAH emission features at
3.3 and 11.3 $\mu$m are weak compared with the emission features located at
6.2, 7.7, and 8.6 $\mu$m, which is also indicating a high degree of ionization
in the population of PAHs (see Figure 2 in Allamandola, Hudgins \& Sandford
1999). Note, however, that if the UV radiation field becomes too strong the PAH
molecules can be destroyed, especially the small ones with a size $\sim$20$-$30
carbon atoms (see e.g. Allain, Leach \& Sedlmayr 1996). This means that the
C-rich dust seen in the ISO spectrum subject to the UV irradiation coming from
the central star must be shielded from the stronger ISM UV radiation field by
the outer layers of the circumstellar shell, where the OH maser emission is
originated. 

For the cool dust emitting at 110 K, a different dust model was assumed,
composed mainly of astronomical silicates. This choice takes into account the
O-rich nature of the crystalline silicates detected in the ISO spectrum at
wavelengths longer than 25 $\mu$m. The crystalline silicates are expected to be
formed in the circumstellar dust shells of evolved stars at temperatures in the
range 60$-$160 K (Molster et al. 2002b). The mean emissivity value adopted
between 25 and 60 $\mu$m was taken from Figure 5 in Draine \& Lee (1984). 
If we try to confine this cool O-rich dust to the observed extension 
of 0.3 $\arcsec$ we would need to adopt a very large dust grain size of $>$0.1
$\mu$m. Note that for this O-rich cool component we derive dust equilibrium
radii of $\sim$0.02, $\sim$0.05 and 0.17 pc (or 1.1$\arcsec$, 3.4$\arcsec$ and
10.4$\arcsec$ at 3.2 kpc) for dust grain sizes of 0.1, 0.01 and 0.001 $\mu$m,
respectively, which in all cases are inconsistent with our mid-infrared
observations. These calculations indicate that independent of the dust grain
size considered, the O-rich cool dust must be located much farther away from
the central source than the C-rich warm dust emission (at 430 K) detected in
the mid-infrared This different relative distribution of O-rich and
C-rich dust would also be consistent with the detection of OH maser emission
from the outer shell and suggests that the material expelled by the central
star during the previous AGB phase was predominantly O-rich.

\subsubsection{Evolutionary Status of IRAS 07027$-$7934}
At present, the evolutionary status of I07027 is not well understood. The
hydrogen-deficiency of the central star together with the mixed dust chemistry
(C-rich and O-rich) is a common finding among the limited sample of known [WC]
PNe (De Marco \& Soker 2002; Cohen et al. 2002). The most promising scenarios
to explain the current observational properties of this rare class of PNe are:
(i) the so-called `disk-storage' scenario (Jura, Chen \& Plavchan 2002;
Yamamura et al. 2000); (ii) a final thermal pulse while the star was still in
the AGB; or (iii) a late thermal pulse during the post-AGB evolution (Herwig
et al. 1997, 1999; Herwig 2000, 2001; Bl\"ocker 2001).

The disk-storage scenario invokes the presence of a binary system in which the
O-rich silicates are trapped in a disk formed by a past mass transfer event,
with the C-rich particles being more widely distributed in the nebula as a
result of recent ejections of C-rich material. This type of dusty disk
structures have been detected in some PPN/PNe with binary [WC] central stars
like CPD$-$56$\degree$8032 (De Marco, Barlow \& Cohen 2002) or in the Red
Rectangle (HD 44179) (Waters et al. 1998), but no firm evidence of the
presence of any disk-like structure nor of the binarity of I07027 exists yet.

Both a final thermal pulse in the AGB and a late thermal pulse during 
the post-AGB phase can eventually produce a sudden switch to a C-rich 
chemistry and a strong stellar wind, which is also characteristic of these 
[WC] CSPNe. However, because of the short lifetime of stars in the post-AGB
phase, the latter is expected to be a rare phenomenon. Models predict that 
post-AGB stars which experience a late thermal pulse evolve back into the AGB
(the so-called ``\textit{born-again}'' scenario, e.g., Herwig 2001; Bl\"ocker
2001). As a result of this, they show a fast spectroscopic evolution in the H-R
diagram as well as peculiar spectroscopic features (e.g., Asplund et al. 1999;
Lechner \& Kimeswenger 2004; Hajduk et al. 2005) which are not observed in
I07027, nor in any other known [WC] CSPNe.

A final thermal pulse in the AGB phase seems to be a more plausible 
explanation since it does not require the assumption of exotic scenarios. As
we have discussed in Section 5.2.4, the emission detected in our mid-infrared
images can be mainly attributed to ionized PAHs plus thermal emission from
relatively warm dust ($\sim$430 K) located very close to the central source.
The OH maser emission detected by Zijlstra et al. (1991) supports the idea that
the envelope of I07027 was until very recently O-rich. It is very difficult to
explain how a low-mass disk around a binary system, which could act as an
oxygen-rich reservoir, may be able to sustain such a luminous maser emission.
Attending to geometry considerations, Zijlstra et al. (1991) suggests that the
star must have changed its chemistry within the last 500 yrs. I07027 may have
experienced a final thermal pulse in the AGB which has produced the recent
switch to a C-rich chemistry. All C-rich material would then be warm as a 
consequence of its very recent formation and, thus, located very close to the
central source (as it is actually observed) while the cooler O-rich material
ejected during the previous AGB phase is then found now only farther away from
the central source.

In contrast, the typical disk sources with dual chemistry which are known to be
binary systems show a completely different relative distribution of O-rich and
C-rich dust. Waters et al. (1998) found that the PAH emission at 11.3 $\mu$m
has a clumpy nature and comes from the extended nebula around HD 44179, while
the O-rich material is located in a circumbinary disk. More recently, the
bipolar post-AGB star IRAS 16279$-$4757 has been studied in the mid-infrared by
Matsuura et al. (2004). They found that the PAH emission is enhanced at the
outflow, while the continuum emission is located towards the center. Thus, they
suggest the presence of a dense O-rich torus around an inner, low density
C-rich region and a C-rich bipolar outflow resembling the morphology attributed
to HD 44179. The observational characteristics of I07027 indicate a totally
different formation mechanism, which are only consistent with a very recent
change of chemistry from O-rich to C-rich. 

\section{Conclusions}
We have presented diffraction limited mid-infrared images of the PPN I16594 and
the [WC] PN I07027 at 8.6, 11.5 and 11.7 $\mu$m taken under exceptionally good
seeing conditions ($\leq$0.5\arcsec). By applying the Lucy-Richardson
deconvolution algorithm, we have resolved, for the first time, the subarcsecond
dust shell structures around both objects.

I16594 displays two emission peaks in the innermost region of the circumstellar
dust shell at the three wavelengths observed. This two-peaked mid-infrared
morphology is interpreted as an equatorial density enhancement 
revealing the presence of a dusty toroidal structure with a 0.4$\arcsec$
radius size (or $\sim$640 AU corresponding to a dynamical age of $\sim$190 yr
at the assumed distance of 1.6 kpc). The observed size is used to
derive the dust temperature at the inner radius of the shell. This result has
been combined with the information derived from the ISO observations of I16594
to conclude that the mid-infrared emission detected in our TIMMI2 images must
be dominated by PAH molecules or clusters which must be mainly distributed
along the torus, as suggested by the similar size and morphology observed in
all filters. We have also found that the axis of symmetry observed in the
mid-infrared is well aligned with only one  of the bipolar outflows (at
P.A.$\sim$84$\degree$) seen as optical reflection nebulae in the optical HST
images. We suggest that the multiple outflow formation has not been coeval and
that, at present, the outflow material is being ejected in this direction. 
Consistently, the H$_2$ shocked-emission seen in the HST NICMOS image is mainly
distributed along the same bipolar axis where the fast post-AGB wind is
interacting with the slow moving material ejected during the previous AGB
phase. The presence of several other bipolar outflows at a variety of position
angles may be the result of past episodic mass loss events. 

I07027 exhibits a slightly asymmetric mid-IR emission core which is only
marginally extended along the north-south direction with FWHM=0.3$\arcsec$ at
8.6 and 11.5 $\mu$m. This is the same orientation observed in recent HST
images of the source taken in the near-infrared. The mid-infrared emission is
attributed to a combination of emission from highly ionized, small PAH
molecules plus relatively warm dust continuum located very close to the central
star. The characteristics of the PAH emission observed in the ISO 
spectrum are also consistent with this interpretation. Taking into account
the spatial distribution of the C-rich material deduced from our observations
and because the OH maser emission from I07027 is expected to be located in the
external and cooler regions, we propose that the dual chemistry observed in
I07027 must be interpreted as the consequence of a recent thermal pulse 
(probably at the end of the previous AGB phase) which has switched the
chemistry of the central star from the original O-rich composition to a C-rich
one within the last 500 yrs. This might be the commom mechanism which
originates the dual chemistry and strong stellar winds usually observed in
other [WC]-type CSPNe.



\acknowledgments
DAGH is grateful to Eva Villaver for her useful comments. AM and PGL
acknowledge support from grants AYA 2001$-$1658 and  AYA 2003$-$9499,
respectively, from the  Spanish Ministerio de Ciencia y Tecnolog\'{\i}a 
(MCYT).

\clearpage

\begin{deluxetable}{ccccccccc}
\tabletypesize{\scriptsize}
\tablecaption{Summary of the ESO 3.6m/TIMMI2 Mid-Infrared Observations.
\label{tbl-1}}
\tablewidth{0pt}
\tablehead{
\colhead{Object} & \colhead{Date} &
\colhead{Filter} &
\colhead{$\lambda_{c}(\Delta\lambda)$}  & \colhead{$T_{Total}$} &
\colhead{Size$^{a}$} &
\colhead{PSF Size$^{b}$}  &
\colhead{Flux$^{c}$}   & \colhead{Peak}\\
\colhead{} & \colhead{} & \colhead{} & \colhead{($\mu$m)}
&\colhead{(s)} &\colhead{(arcsec)} &\colhead{(arcsec)} &\colhead{(Jy)} &
\colhead{(Jy arcsec$^{-2}$)} 
}
\startdata
IRAS 16594$-$4656&2001 Oct 9&N1&8.6(1.67)&120.4&$\sim$1.7 x
1.6&0.57&14&5\\
$\dots$&$\dots$&N11.9&11.5(1.89)&107.5&$\sim$1.8 $\times$
1.6&0.73&40&13\\
$\dots$&$\dots$&SiC&11.7(3.21)&161.3&$\sim$1.8 $\times$ 1.6&0.75&35&11\\
IRAS 07027$-$7934&2001 Oct 10 &N1&8.6(1.67)&120.4&0.65&0.57&19&23\\
$\dots$&$\dots$&N11.9&11.5(1.89)&107.5&0.83&0.73&19&17\\
 \enddata


\tablenotetext{a}{Major and minor axis length at 40\% of the peak intensity or
FWHM in the case of IRAS 07027$-$7934}
\tablenotetext{b}{FWHM of the mean PSF adopted}
\tablenotetext{c}{The fluxes are not color corrected}
\end{deluxetable}

\clearpage

\begin{deluxetable}{cccccccccc}
\tabletypesize{\scriptsize}
\tablecaption{Mid-infrared Toroidal PPN/PNe sample. \label{tbl-2}}
\tablewidth{0pt}
\tablehead{
\colhead{IRAS Name} & 
\colhead{Torus$^{a}$}& \colhead{Ref.$^{b}$}  & \colhead{Sp.Type$^{c}$} &
\colhead{Chem.$^{d}$}  &
\colhead{Type$^{e}$} & \colhead{Morph.$^{f}$} & \colhead{PAHs$^{g}$}
&\colhead{21 $\mu$m$^{h}$}&
\colhead{Ref.$^{i}$}  
}
\startdata
07134$+$1005 &  R   & 1& F5 Iab & C & PPN & S+B&y&y&1\\
16594$-$4656 &  R   & 2& B7    & C & PPN & S+M&y&y&2\\
17436$+$5003 &  R   & 3& F3 Ib  & O & PPN & S+B&n&n&3\\
19114$+$0002 &  R$^{*}$   & 4& G5 Ia  & O & PPN/SG & S+M&n&n&3
\\
21282$+$5050 &  R   & 1& O9& C & Young PN & S+B&y&n&4 \\
22223$+$4327 &  U   & 4& G0 Ia  & C & PPN & S+M &y&y&5 \\
22272$+$5435 &  R   & 5& G5    & C & PPN & S+M&y&y&6 \\
\enddata
\tablenotetext{a}{R: Two emission peaks clearly resolved. R$^{*}$: The torus
orientation is almost pole-on and it is seen as a ring. U: The two emission
peaks are not resolved but there are evidences for the presence of a toroidal
structure.} 
\tablenotetext{b}{References for the mid-infrared images and their
classification: 1) Meixner et al. (1997); 2) This work; 3) Ueta, Meixner \&
Bobrowsky (2000); 4) Meixner et al. (1999); 5) Ueta et al. (2001).}
\tablenotetext{c}{Spectral types taken from SIMBAD.}
\tablenotetext{d}{C: C-rich. O: O-rich.}
\tablenotetext{e}{Evolutionary classification.}
\tablenotetext{f}{S+B: Star + bipolar. S+M: Star + multipolar. Optical
morphology taken from Meixner et al. (1999). For IRAS 22223$+$4327 and IRAS
16594$-$4656 the HST images were retrieved from the HST Data Archive.}
\tablenotetext{g}{Presence of the dust features at 3.3, 6.2, 7.7, 8.6 and 11.3
$\mu$m generally attributed to PAHs.}
\tablenotetext{h}{Presence of the unidentified dust emission feature at 21
$\mu$m.}
\tablenotetext{i}{References for the mid-infrared spectra: 1) Hrivnak, Volk \&
Kwok (2000); 2) Garc\'{\i}a-Lario et al. (1999); 3) Molster et al. (2002a); 4)
Justtanont et al. (1996); 5) Kwok, Hrivnak \& Geballe (1995); 6) Hrivnak, Volk
\& Kwok (1999).}
\end{deluxetable}




\clearpage

\begin{figure}
\epsscale{.80}
\plotone{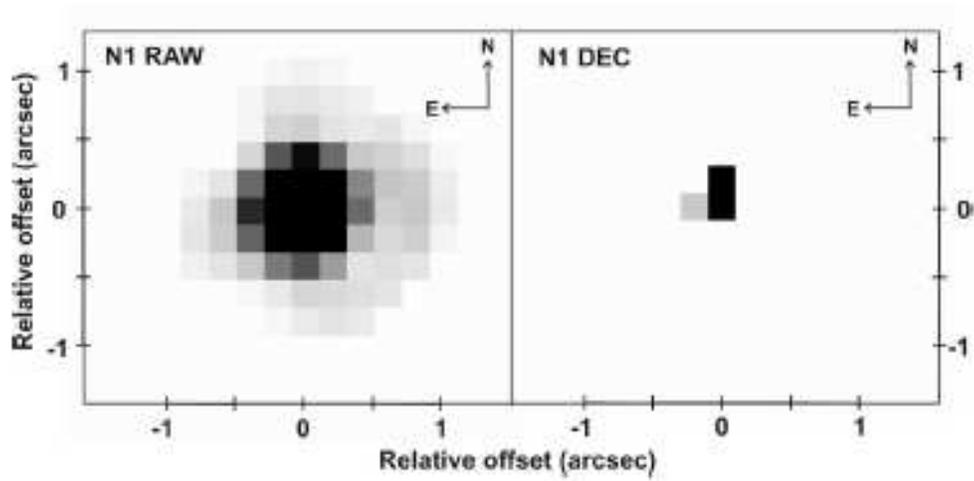}
\caption{Illustrative example of the deconvolution of one of the observed 
standard stars showing the goodness of the deconvolution process. Raw 
mid-infrared image in the N1 (8.6 $\mu$m) filter (left) and the corresponding
deconvolved image using the Lucy-Richardson algorithm (right).\label{fig1}}
\end{figure}

\clearpage

\begin{figure}
\epsscale{.70}
\plotone{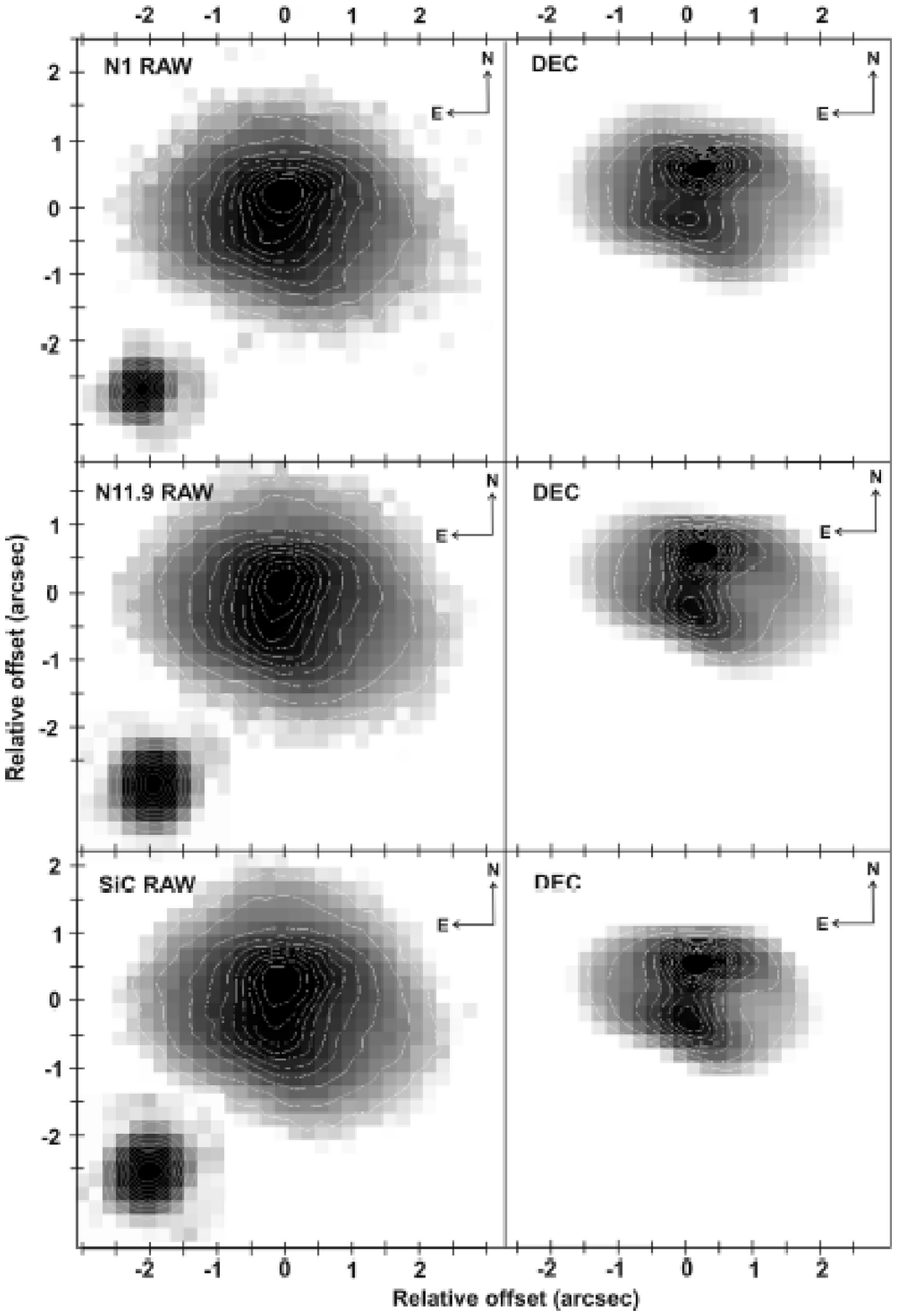}
\caption{From top to bottom, raw mid-infrared images of IRAS 16594$-$4656 in
the N1 (8.6 $\mu$m), N11.9 (11.5 $\mu$m) and SiC (11.7 $\mu$m) filters (left)
and the corresponding deconvolved images using the Lucy-Richardson algorithm
(right). The tick marks show relative offsets from the center of the nebula in
arcseconds. Contours range from 10\% to 90\% of the peak intensity (in steps of
10\%) plus the outermost contour, which corresponds to 5\% of the peak
intensity (or $\sim$60 $mJy$ at 10-$\sigma$ level above the sky background).
The insets show the standard star PSFs in each filter.\label{fig2}}
\end{figure}

\clearpage 

\begin{figure}
\includegraphics[angle=-90,scale=.50]{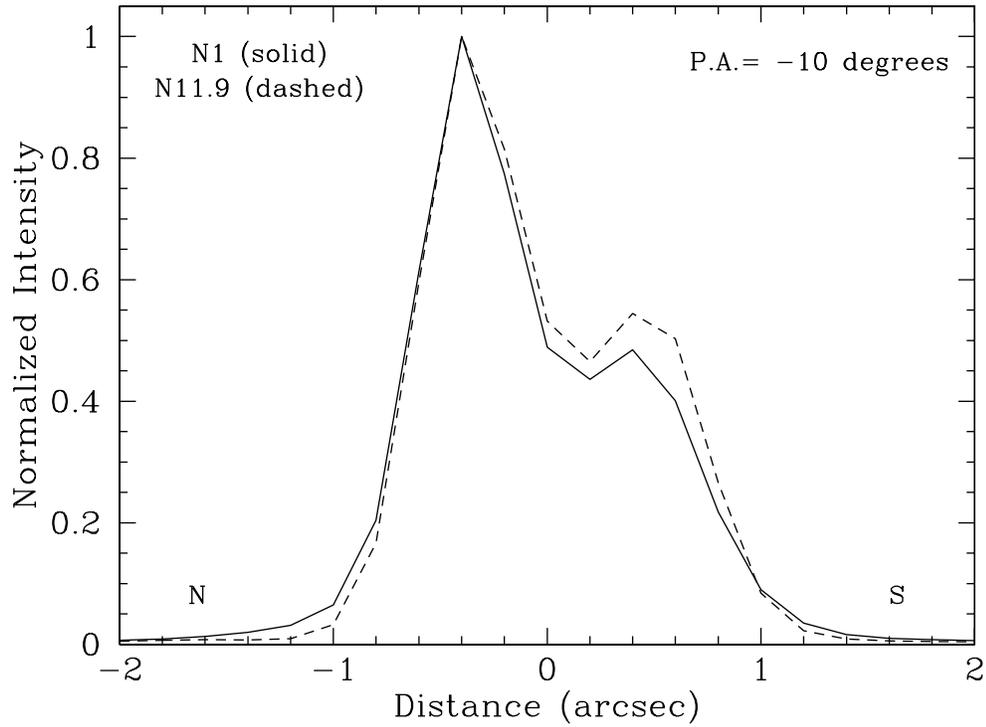}
\caption{Normalized surface intensity profiles along the dust torus direction
(from north to south) in IRAS 16594$-$4656. The solid and dashed lines
correspond to the N1 (8.6 $\mu$m) and N11.9 (11.5 $\mu$m) filters,
respectively. The center of the nebula (tentatively identified as the central
star position, see text) corresponds to distance 0. \label{fig3}}
\end{figure}

\clearpage 
\begin{figure}
\epsscale{.80}
\plotone{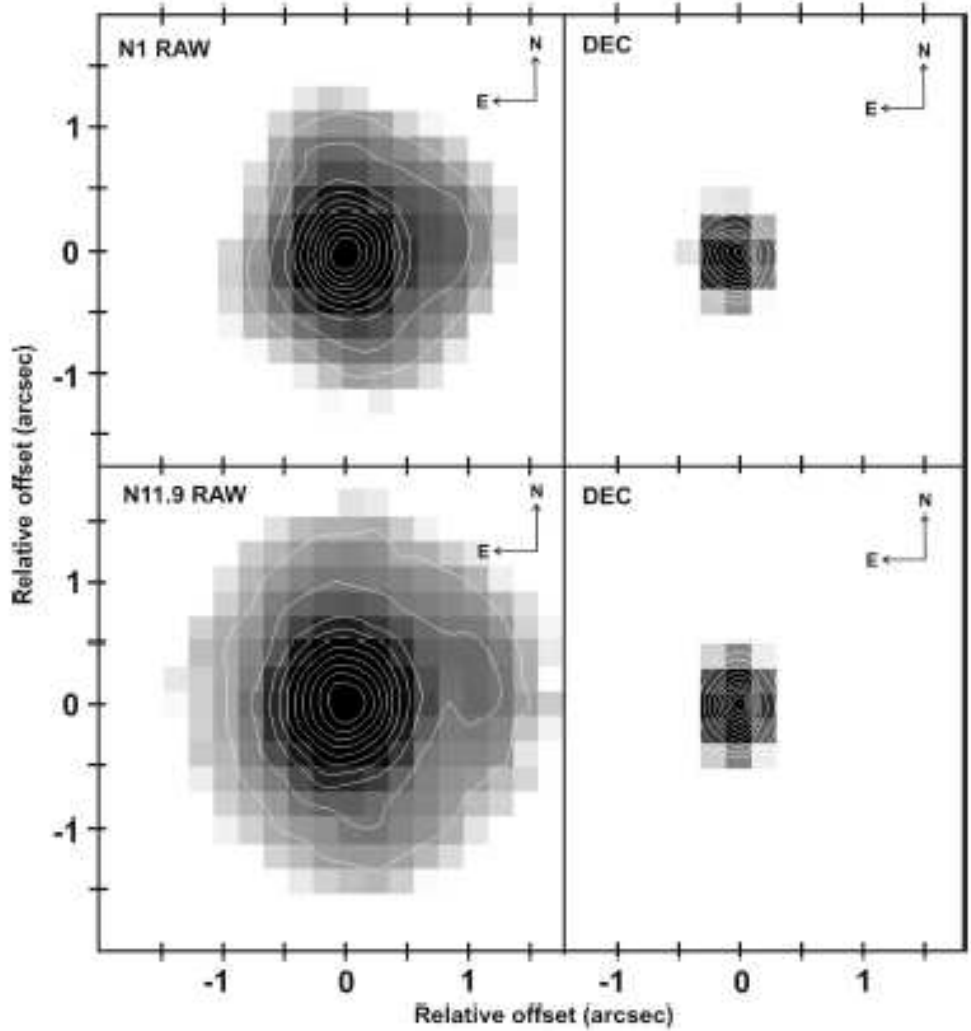}
\caption{From top to bottom, raw mid-infrared images of IRAS 07027$-$7934 in
the N1 (8.6 $\mu$m), and N11.9 (11.5 $\mu$m) filters (left) and the
corresponding deconvolved images using the Lucy-Richardson algorithm (right).
The tick marks show relative offsets from the center of the nebula in
arcseconds. Contours range from 10\% to 90\% of the peak intensity (in steps of
10\%) plus the outermost contour, which corresponds to 5\% of the peak
intensity (or $\sim$50 $mJy$ at 7-$\sigma$ level above the sky background).
\label{fig4}}
\end{figure}

\clearpage 

\begin{figure}
\plotone{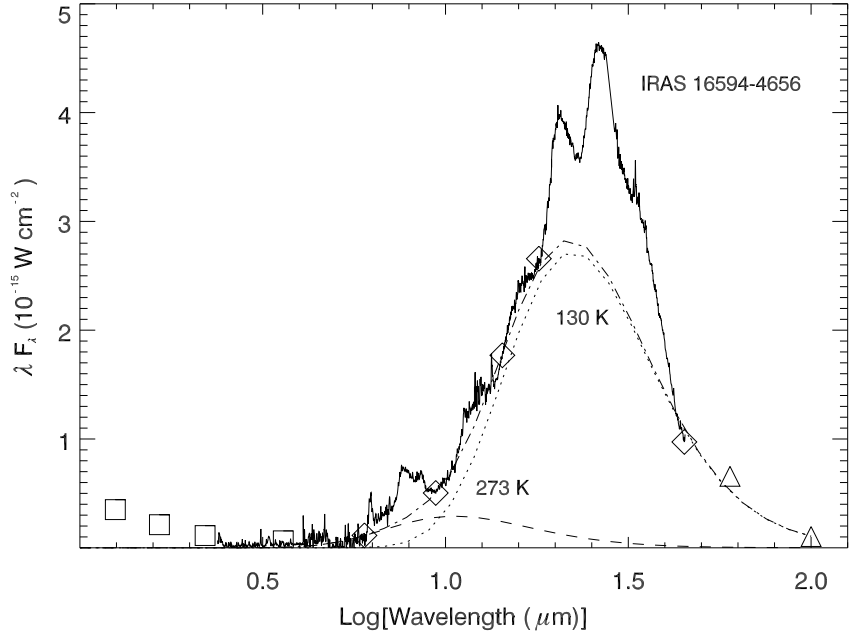}
\caption{Two blackbody fitting of the SED of IRAS 16594$-$4656. The square
symbols correspond to the dereddened JHKL photometry while the continuum points
at 6, 9.4, 14.3, 18, and 45 $\mu$m that were carefully selected for the fitting
from the SWS ISO spectrum correspond to the diamond symbols. The IRAS
photometry at 60 and 100 $\mu$m are represented with triangle symbols. The best
fit (dotted and dashed line) is obtained with the combination of two
blackbodies with temperatures of 273 K (dashed line) and 130 K (dotted line),
respectively. Note that the huge excess emission between 18$-$45 $\mu$m is
mainly due to the presence of the strong 21 $\mu$m feature, the 26 $\mu$m
feature and the broad 30 $\mu$m feature (see text).\label{fig5}}
\end{figure}

\clearpage 

\begin{figure}
\plotone{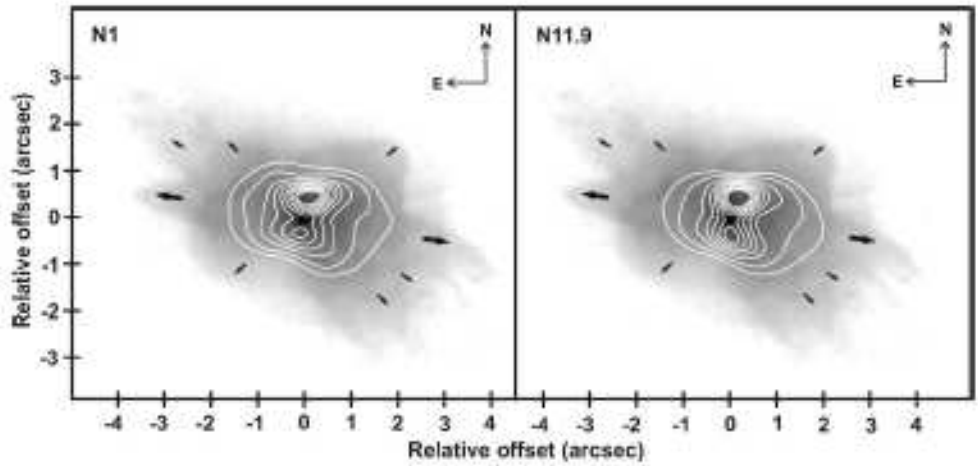}
\caption{Contour maps of the deconvolved mid-infrared images taken in the N1
(8.6 $\mu$m) and N11.9 (11.5 $\mu$m) filters overlaid on the optical HST image
of IRAS 16594$-$4656 in the F606W filter (taken from the HST Data Archive). The
tick marks show relative offsets from the center of the nebula in arcseconds.
The arrows indicate the direction of the various bipolar lobes identified in
the optical images. The big one marks the lobe whose orientation is remarkably
coincident with the mid-infrared axis defined by the dust torus.\label{fig6}}
\end{figure}

\clearpage 

\begin{figure}
\plotone{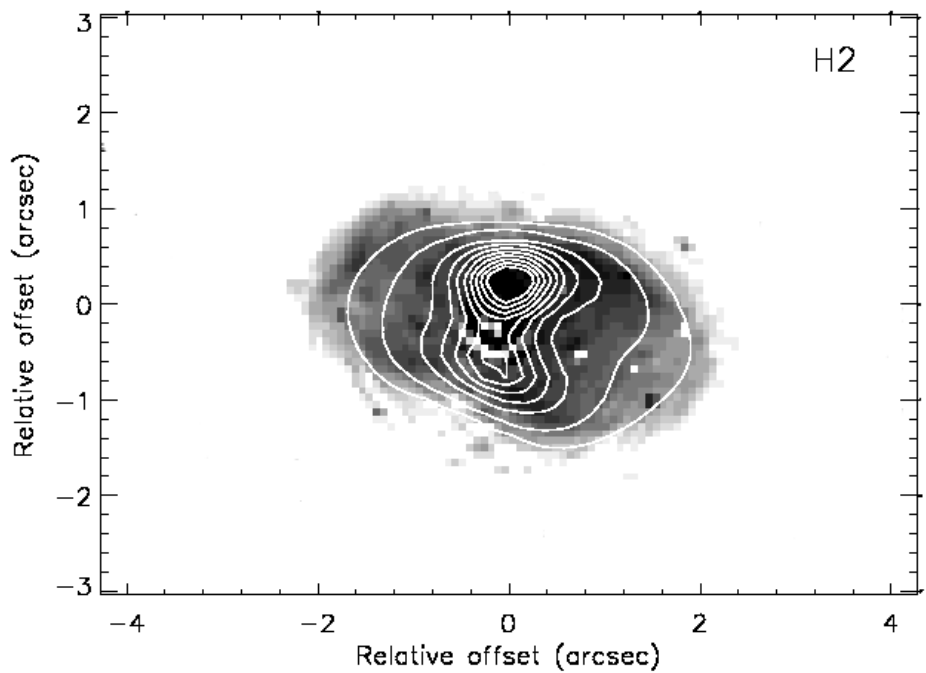}
\caption{Contour map of the deconvolved mid-infrared image taken in the N11.9
(11.5 $\mu$m) filter overlaid on the HST-NICMOS image (taken from the HST Data
Archive) showing the near-infrared H$_2$ emission at 2.122 $\mu$m detected in
IRAS 16594$-$4656. The orientation is the same as in the other figures of the
paper (north is up and east to the left) and the H$_2$ image is displayed on
logarithmic-scale. Remarkably, the H$_2$ emission which nicely coincides with
the mid-infrared emission is mainly distributed along the walls of the lobe at
P.A.$\sim$84$\degree$ identified in the optical images of IRAS 16594$-$4656
(see Fig. 6). Note that the H$_2$ image was continuum-subtracted using the
HST-NICMOS image containing the emission in the adjacent continuum at 2.15
$\mu$m. \label{fig7}}
\end{figure}

\clearpage 

\begin{figure}
\plotone{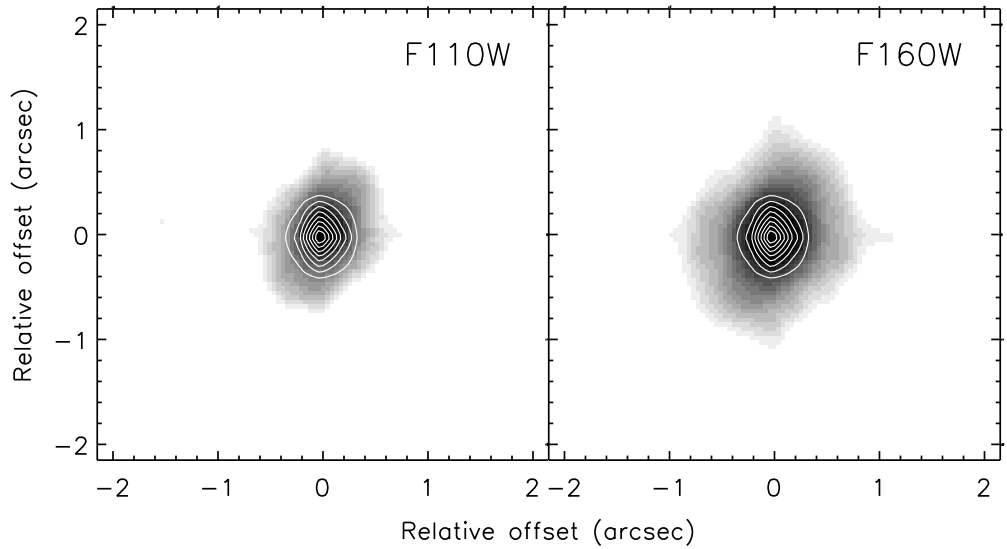}
\caption{Contour map of the deconvolved mid-infrared image taken in the N11.9
(11.5 $\mu$m) filter overlaid on the logarithmic-scale near-infrared HST-NICMOS
images of IRAS 07027$-$7934 (taken from the HST Data Archive) in the F110W
(left panel) and F160W filters (right panel). Again, north is up and east to
the left. Note that the near-infrared images show that indeed the inner regions
have an elongated shape along the north-south direction, coincident with the
orientation observed in our mid-infrared images (see text for more details).
\label{fig8}}
\end{figure}

\end{document}